# Anticipated emotions associated with trust in autonomous vehicles

Lilit Avetisian, Jackie Ayoub, Feng Zhou
University of Michigan Dearborn

Trust in automation has been mainly studied in the cognitive perspective, though some researchers have shown that trust is also influenced by emotion. Therefore, it is essential to investigate the relationships between emotions and trust. In this study, we explored the pattern of 19 anticipated emotions associated with two levels of trust (i.e., low vs. high levels of trust) elicited from two levels of autonomous vehicles (AVs) performance (i.e., failure and non-failure) from 105 participants from Amazon Mechanical Turk (AMT). Trust was assessed at three layers i.e., dispositional, initial learned, and situational trust. The study was designed to measure how emotions are affected with low and high levels of trust. Situational trust was significantly correlated with emotions that a high level of trust significantly improved participants' positive emotions, and vice versa. We also identified the underlying factors of emotions associated with situational trust. Our results offered important implications on anticipated emotions associated with trust in AVs.

## INTRODUCTION

Automated driving is the next evolutionary step in mobility with main benefits of increasing safety, efficiency, and comfort. Trust plays a pivotal role in the acceptance and adoption of autonomous vehicles (AVs) (Ayoub et al., 2019) (Ayoub et al., 2021) (Ayoub & Zhou, 2020) (Ayoub et al., 2021). Trust can be defined as "the attitude that an agent will help achieve in a situation characterized by uncertainty and vulnerability" (Lee and See 2004). Therefore, promoting an appropriate level of trust is essential for interacting with AVs, which can be realized by trust calibration. An appropriate calibration process can avoid over-trust and under-trust, which in turn avoid misuse and disuse of the system, respectively (Parasuraman & Riley, 1997) (Lee & See, 2004).

Despite the importance of trust, fundamental questions remain unanswered about how trust actually evolves in reality, which makes it difficult to calibrate trust during the human-machine interaction process. Researchers have investigated the effects of different factors on trust from the cognitive perspective. For instance, Beggiato et al. (2015) and Ebnali et al. (2020) worked on building an appropriate mental model of AV trust through training. Beller at al. (Beller et al., 2013) proposed a real-time feedback of system information (i.e., uncertainty, safety, and performance) to calibrate trust. Koo et al. (2015) and Du et al. (2019) studied the effects of explaining AVs' upcoming actions and error information on trust.

In addition to the cognitive perspective, researchers have identified that trust is also influenced by emotions (Lee & See, 2004) (Madsen & Gregor, 2000). Dunn and Schweitzer (2005) showed that important trust decisions were made in contexts influenced by affect. This is especially true when the decision is emotion-rich (e.g., emotion significantly affected drivers' takeover readiness and performance in highly automated driving (Du et al., 2019)).

However, the effect of emotions on human-automation trust has been largely overlooked by researchers. Thus, understanding how emotions relate to trust is essential to human-machine interaction. Researchers have examined the relation between emotion and trust in different ways. Boone and Buck (2003) showed that being emotionally expressive is a marker of trustworthiness. For instance, facial expressions and body language can be used to track trustworthiness. Lee and Selart (Lee & Selart, 2012) provided an evaluation on the Autobiographical Emotional Memory Task which was a manipulation technique of emotions. Their results showed that negative emotions decreased trust, but only if those negative emotions produced low certainty appraisals. Researchers also developed models, such as affect as information (Schwarz & Clore, 1983) and affect infusion (Forgas, 1995). The first model suggests that people use their mood to guide their judgment. The latter model suggests that emotions influenced initial trust judgments.

Although such models provide better understanding of how emotions influence people's judgements in general, the relationship between emotions and trust in automated driving is still not known. Thus, we conducted a study to investigate the structure of anticipated emotions associated with low and high levels of trust related to different autonomous vehicles (AVs) performances (i.e., failure and non-failure).

## METHOD

**Participants**

A total number of 121 participants were recruited from Amazon Mechanical Turk (AMT) and completed an online survey. The study was designed for participants who were aged 18 years or older and located in the United States. All the participants had a valid US driver's license, completed at least 1000 Human Intelligence Task with 95% approval rate. The research was approved by the Institutional Review Board at the University of Michigan (i.e., Application number is HUM00200349). In addition, attention questions were included in the survey to exclude participants who provided nonsensical results. After the screening, 105 participants' responses (44 females and 61 males; M = 37.0 years and SD =

11.1 years) were included for analysis. Participants were randomly assigned to the experiment conditions (i.e., failure, non-failure) and were compensated with $2.5 after completing the survey. On average, the survey took about 17 min to complete the survey.

**Apparatus**

An online survey was conducted using Amazon Mechanical Turks (AMT) (Seattle, WA, www.mturk.com/). AMT is a web-based survey company, operated by Amazon Web Services (Paolacci et al., 2010). The questionnaire was developed using a Qualtrics (Provo, UT, www.qualtrics.com/) online tool and was published in AMT. Responses approval and participant's payments were managed using AMT.

**Experiment design**

The purpose of the study is to investigate the effect of participants' trust (i.e., low vs high levels of trust) on their anticipated emotions under different system performances (i.e., failure and non-failure). The independent variable was the trust level of the participants (i.e., low vs high) elicited from different system performance (i.e., failure and non-failure). Trust was assessed at three layers (i.e., dispositional, initial learned, and situational trust) proposed by (Hoff and Bashir, 2015). To explore participants' overall tendency to trust AVs, dispositional trust was assessed with the six items scale proposed by (Merritt 2011). To investigate participants' tendency to trust AVs based on past/current experience and prior to any interaction with the AV system, Manchon et al. (2021)'s ten-item questionnaire was used to measure Initial Level of Trust in Automated Driving (TiAD). To measure participants' tendency to trust AVs based on the situation, the Situational Trust Scale for Automated Driving (STS-AD) that included six items (Holthausen et al., 2020) was used. All three layers of trust were measured using a 7-point Likert scale. As for the dependent variable, participants' emotions were assessed using subjective ratings of a list of 19 emotion words (i.e., Disdainful, scornful, contemptuous, hostile, resentful, ashamed, humiliated, confident, secure, grateful, happy, respectful, nervous, anxious, confused, afraid, freaked out, lonely, isolated) (Jensen et al., 2020). These discrete emotions can capture both analytical and syncretic-affective knowledge developed from affective neuroscience (Chaudhuri 2006). They have also been used to investigate emotions in decision making and trust in human-machine automation (Buck et al. 2018).

**Procedure**

First, participants were asked to sign a consent form that they agreed to participate in the study. Following the consent, participants answered the first attention question about AVs' challenges faced on the roads. Second, the participants filled out a list of demographic questions. The third section included participants' dispositional trust assessment. After this section, participants evaluated their emotions regarding AVs using 19 emotion items without receiving any information regarding AVs' performance. In the fourth section, participants needed to read a text paragraph, presenting basic information related to AVs (see Table 1). For instance, in the high trust condition, the presented information showed positive facts related to AVs. Whereas in the low trust condition, the presented information showed negative facts related to AVs. Based on the presented information, participants were asked to rate their initial learned trust. After that, the participants went through another attention question related to the challenges faced by AVs if the tested condition was the low trust. If the tested condition was high trust, the attention question was related to the advantages of AVs. In section six, if the tested condition was low trust, the participants watched a video of an AV failing to handle a situation. If the tested condition was high trust, the video was about an AV successfully handling a situation (See Table 1). Then, the participants rated their anticipated emotions to AVs using the 19 emotion items. Next, we evaluated participants' situational trust. At the end of the survey, participants answered a third attention question related to the challenges or the advantages of AVs depending on the tested condition.

Table 1. Description of the advantages and disadvantages of AVs shown in the survey.

| Condition | Textual information and links to videos |
|---|---|
| Low trust | At the moment, self-driving vehicles have a higher rate of accidents compared to human-driven vehicles, but the injuries are less serious. On average, there are 9.1 self-driving vehicle accidents per million miles driven, while the same rate is 4.1 crashes per million miles for human-driven vehicles. Self-driving vehicles had a higher rate of injury per crash: 0.36 injuries per crash, compared with 0.25 for human-driven vehicles (Carsurance, 2022). <br><br> URL: https://tinyurl.com/scfailure |
| High trust | The safety benefits of self-driving vehicles are paramount. Self-driving vehicles' potential to save lives and reduce injuries is rooted in one critical and tragic fact: 94% of serious crashes are due to human error. Self-driving vehicles have the potential to remove human error from the crash equation, which will help protect drivers and passengers, as well as bicyclists and pedestrians. When you consider more than 35,000 people die in motor vehicle-related crashes in the United States each year, you begin to grasp the lifesaving benefits of driver assistance technologies. (Automated Vehicles for Safety ) <br><br> URL: https://tinyurl.com/nonfailure |

**Scenarios eliciting trust levels**

As mentioned previously, two videos were used to evaluate situational trust in AVs, as suggested by Holthausen et al. (2020). Previous studies showed that videos were one of the effective ways to ensure higher engagement with information content (Tempelman, 2006). The contents of the experiment scenarios were taken from published real videos on YouTube where an AV was involved in an accident or successfully handled a critical situation. In the high trust condition, the video included 8 short recordings (taken by AVs) where the AVs properly responded to the situation. For example, one of the recordings showed a situation where an incoming vehicle started to dangerously merge into the AV's lane, but the AV quickly responded to the situation and avoided the collision. In the low trust scenario, the video showed an AV failed to detect an overturned semi-truck on the road and crashed. The original audios were replaced with explanations/descriptions of the videos' content recorded by one of the researchers.

**Data analysis**

In this study, we investigated participants' anticipated emotions associated with the two levels of trust (i.e., low vs. high levels of trust). To identify the underlying structure of the emotions associated with trust, we first conducted an exploratory factor analysis (EFA) and reduced the emotion items' dimension into basic components by creating a new low-dimensional subset from the 19 emotion items separately for each experiment condition. The data was analyzed with the R language in the RStudio environment.

## RESULTS

**Manipulation check**

A one-way ANOVA was conducted to investigate the effect of AV performance (i.e., failure, non-failure) on trust measures. Results showed that the effect of AV performance was not significant on the dispositional trust ($F(1,104) = .143$, $p = .705$). With regard to initial learned trust, no significant effect was found ($F(1,104) = .629$, $p = .429$). On the other hand, a significant effect was found on participants' situational trust ($F(1,104) = 19.715$, $p = .000$). In the high trust condition, participants' ratings for trust were significantly higher than those in the low trust condition.

**Comparisons of emotions in two levels of trust**

A comparison of the emotion items showed that there was no significant difference between the two tested conditions (i.e., low vs high levels of trust) before watching the videos ($p > .05$ for all emotion items). However, a significant effect was observed after watching the videos (see Table 2). In the low trust condition, all the negative emotion ratings were significantly higher compared to the high trust condition ($p < .05$). Furthermore, significantly higher ratings of positive emotions were found in the high trust condition ($p < .05$). Such results indicated that a high level of situational trust increased participants' positive emotions and reduced their negative emotions, and vice versa.

**Underlying factors of emotions in different levels of trust**

Table 2. Difference of individual emotions after watching the videos

| Emotions | Low Trust Mean ± SD | High Trust Mean ± SD | F | p |
|---|---|---|---|---|
| Disdainful | 4.353 ± 1.730 | 3.036 ± 2.108 | 12.245 | 0.000 |
| Scornful | 4.686 ± 1.655 | 3.182 ± 2.038 | 17.247 | 0.000 |
| Contemptuous | 4.843 ± 1.528 | 3.273 ± 1.976 | 20.714 | 0.000 |
| Hostile | 4.902 ± 1.628 | 3.527 ± 2.116 | 13.898 | 0.000 |
| Resentful | 4.863 ± 1.721 | 3.364 ± 2.040 | 16.592 | 0.000 |
| Ashamed | 4.784 ± 1.911 | 3.491 ± 2.210 | 10.314 | 0.001 |
| Humiliated | 4.882 ± 2.016 | 3.989 ± 2.297 | 4.571 | 0.035 |
| Confident | 4.039 ± 2.010 | 5.055 ± 1.660 | 8.088 | 0.005 |
| Secure | 3.804 ± 2.030 | 4.909 ± 1.567 | 9.928 | 0.002 |
| Grateful | 4.000 ± 2.059 | 5.164 ± 1.463 | 11.377 | 0.001 |
| Happy | 4.078 ± 2.189 | 4.836 ± 1.607 | 4.169 | 0.044 |
| Respectful | 4.118 ± 2.160 | 5.915 ± 0.740 | 4.959 | 0.028 |
| Nervous | 5.686 ± 1.476 | 4.418 ± 1.863 | 14.933 | 0.000 |
| Anxious | 5.569 ± 1.487 | 4.382 ± 2.095 | 11.155 | 0.001 |
| Confused | 4.922 ± 1.730 | 3.473 ± 1.824 | 17.538 | 0.000 |
| Afraid | 5.569 ± 1.360 | 3.636 ± 1.850 | 37.063 | 0.000 |
| Freaked out | 5.490 ± 1.223 | 4.055 ± 1.830 | 22.194 | 0.000 |
| Lonely | 4.431 ± 2.042 | 3.600 ± 1.978 | 4.532 | 0.036 |
| Isolated | 4.588 ± 2.080 | 3.636 ± 2.058 | 5.602 | 0.019 |

We conducted EFA for before (B) and after (A) the situational trust manipulation in both low trust and high trust conditions and identified the latent factors of emotions, which were sufficient to explain the underlying structure of emotions. Since there was no significant difference between in trust before the participants watched the video. We emphasized the emotion structure after they watched the video.

In the low trust condition, the EFA results showed that three factors explained 75.3% variability of emotions in the B state. Although the Chi-square test indicated that three factors were not sufficient and more factors were needed to explain the variability ($p = .001$), the subsets with four or five factors did not show significant influence. Based on the normalized eigenvalues, the first factor explained 45.0% of the variance. Here, 12 emotion items (i.e., Disdainful, Scornful, Contemptuous, Hostile, Resentful, Ashamed, Humiliated, Confused, Afraid, Freaked Out, Lonely, and Isolated) showed higher than .70 factor loadings. The second factor explained 19.4% of the variance and represented 5 positive emotions (i.e., Confident, Grateful, Secure, Happy, Respectful) with .70 and higher factor loadings. The third factor explained 10.8% of the variance from the emotions and included 2 negative items (i.e., Nervous, Anxious) with .60 and higher factor loadings.

For the A state, a four-factor model was sufficient to explain the 70.6% variability as shown in Table 3. The first factor (Resentfully Aversion) included 6 emotions (i.e., Contemptuous, Hostile, Scornful, Disdainful, Resentful, Humiliated) with .50 and higher factor loadings and explained 21.1%. The second factor (Happily Acceptance) explained 23.3% of the 5 positive emotions (i.e., Confident, Grateful, Secure, Happy, Respectful) with .76 and higher loadings. The third factor (Nervously Fear) explained 16.6% of the variance with 5 negative emotions (i.e., Confused, Afraid, Freaked out,

Nervous, Anxious) with more than .70 factor loadings. The fourth factor (Lonely Isolated) included 2 negative emotions (i.e., Lonely, Isolated) with more than .60 factor loadings and explained 10.0% variance.

Table 3. After state factor loadings in the low trust condition.

| Emotions | Resentfully Aversion (F1) | Happily Acceptance (F2) | Nervously Fear (F3) | Lonely Isolated (F4) |
|---|---|---|---|---|
| Disdainful | .773 | | | |
| Scornful | .842 | | | |
| Contemptuous | .824 | | | |
| Hostile | .769 | | | |
| Resentful | .640 | | | |
| Ashamed | .486 | | | |
| Humiliated | .509 | | | |
| Confident | | .834 | | |
| Secure | | .845 | | |
| Grateful | | .951 | | |
| Happy | | .878 | | |
| Respectful | | .766 | | |
| Nervous | | | .642 | |
| Anxious | | | .686 | |
| Confused | | | .638 | |
| Afraid | | | .620 | |
| Freaked out | | | .740 | |
| Lonely | | | | .870 |
| Isolated | | | | .680 |

Table 4. After state factor loadings in the high trust condition.

| Emotions | Resentfully Aversion (F1) | Happily Acceptance (F3) | Nervously Fear (F2) |
|---|---|---|---|
| Disdainful | .780 | | |
| Scornful | .748 | | |
| Contemptuous | .756 | | |
| Hostile | .862 | | |
| Resentful | .921 | | |
| Ashamed | .904 | | |
| Humiliated | .769 | | |
| Confident | | .821 | |
| Secure | | .957 | |
| Grateful | | .601 | |
| Happy | | .710 | |
| Respectful | | .536 | |
| Nervous | | | .641 |
| Anxious | | | .728 |
| Confused | | | .593 |
| Afraid | | | .801 |
| Freaked out | | | .658 |
| Lonely | .732 | | |
| Isolated | .756 | | |

In the high trust condition, the EFA led to a four-factor model for B state which contained 74.5% of the total information. The first factor aggregated 30.4% of the variance, explaining 7 negative emotions (i.e., Humiliated, Resentful, Ashamed, Isolated, Hostile, Lonely, Contemptuous) where factor loadings of each emotion item was higher than .62. The second factor explained 17.2% of emotions' variance with more than .60 item loadings of 5 negative emotions (i.e., Confused, Disdainful, Afraid, Scornful, Freaked out). The third factor represented 16.7% of variance of the B state including the positive emotions with .67 and higher loadings (i.e., Confident, Grateful, Secure, Happy, Respectful). The fourth factor explained 10.3% of the variance and included 2 negative items (i.e., Nervous, Anxious) with .73 and .84 factor loadings.

For the A state, three principal factors were extracted according to FA results explaining 32.2%, 18.3 and 15.4 % of the variance. Here, the first factor (Resentfully Aversion) explained the high correlations (.74 or higher) between 9 negative emotion items (i.e., Disdainful, Scornful, Contemptuous, Hostile, Resentful, Ashamed, Humiliated, Lonely, Isolated). The second factor (Happily Acceptance) represented the remaining 5 negative items (i.e., Nervous, Anxious, Confused, Afraid, Freaked out) with .64 or higher loadings. The third factor (Nervously Fear) represented the information of 5 positive emotion items (i.e., Confident, Grateful, Secure, Happy, Respectful), which had more than .53 loading index. Table 3 and Table 4 summarizes the factor loadings of A state with regard to two experiment conditions.

## DISCUSSION AND CONCLUSION

This study aimed to investigate the structure of anticipated emotions associated with trust in autonomous vehicles. Particularly, 19 emotion items were explored to identify the underlying structure of emotions associated with three trust levels via collecting subjective ratings about participants' respective trust and emotions.

Participants did not report significantly different initial trust in the two performance conditions (i.e., failure and non-failure). Particularly, we manipulated participants' initial trust by presenting information about AV's advantages /disadvantages correspondingly and expected that provided information would positively/negatively influence the trust level. This might be due to the fact that we included overall positive (i.e., non-failure) or negative (i.e., failure) information rather than performance-related information so that participants did not feel much difference in initial learned trust in the AV systems. Likewise, there was no significant difference between the anticipated emotions between the two conditions.

On the other hand, our results showed that emotions were significantly influenced by the perceived situational trust. We noticed that in the low trust condition participants had a significant decrease in positive emotions as well as a significant increase in negative emotions compared to those in the high trust condition. This finding was consistent with previous studies (Jensen et. al., 2019) and indicated that a higher level of situational trust led to a higher level of positive emotions and a lower level of negative emotions, and vice versa in the low trust condition.

There was a difference between low trust and high trust conditions with regard to the emotions' underlying structure. The factor analysis showed that in the B state of the low trust condition, almost all negative emotions were correlated and were explained by two factors compared to the A state where participants reported more expressive negative emotions and three factors were needed to explain these emotions. In contrast, in the low trust condition participants had more differentiable negative emotions in B state explained with three factors while in the A state the video manipulation helped to mitigate the intensity of negative emotions and explain them with two factors. In regard to the positive emotions, no difference was found between the B and A states in both conditions. The reason might be that the video elicited a similar structure of positive emotions but in opposite directions respective to the experiment conditions. The failure video that elicited a low level of trust decreased the ratings of all positive emotions while the success video that elicited a high level of trust increased the ratings of all positive emotions leading to the analogical responses in opposite directions. More positive emotion items might be needed to understand the underlying structure better.

Our study also has limitations. First, the scenarios were presented in a low fidelity using YouTube videos, which could potentially influence participants, engagement level and perceived risks. In future studies, the scenarios can be implemented in a driving simulator or in virtual reality. Second, we only collected self-reported measures of trust and emotions, which could be subject to biases. In the future, we plan to collect physiological and behavioral measures which are less susceptible to voluntary control and biases to have a better understanding about the relationships between emotions and trust in automated driving.

As an emotional and cognitive response, once the relationships between the latent structure of emotion in trust-based interaction and trust in AVs are identified, emotion can effectively help build and calibrate driver trust in AVs using affect heuristics.

## ACKNOWLEDGMENT

This work was supported by the University of Michigan-Dearborn Campus Grant and in part by the National Science Foundation.